\magnification=1200
\baselineskip=14pt

\def\as1{\overline{A_{1}}}

\def\bs1{\overline{B_{1}}}

\def\nib{n_{PBH}}
\def\omnor{\omega_{NOR}}

\def\zre{z_{R}}

\def\tri{t_{r}^{-1}}
\def\ppi{t_{pi}^{-1}}
\def\tco{t_{coll}^{-1}}
\def\tee{T_{e}}

\def\part{\partial t}

\def\paro{\partial \omega}

\def\rax{\dot R}
\def\scal{{\rax\over R}}
\def\num{n_{\gamma}}
\def\nure{n_{\gamma R}}
\def\nupi{n_{\gamma PI}}

\def\nugg{n_{\gamma PR}}
\def\parn{\partial n_{\gamma}(\omega,t)}

\def\park{\partial\nugg}
\centerline{\bf REIONIZATION OF THE UNIVERSE}
\centerline{\bf INDUCED BY PRIMORDIAL BLACK HOLES.}
\vskip 1cm
\centerline{\bf{Marina Gibilisco}}\vskip 1mm
\centerline{\it Universit\'a degli Studi di Pavia,}\vskip 1mm
\centerline{\it and INFN sezione di Pavia,}\vskip 1mm
\centerline{\it Via Bassi 6, 20127 Pavia, Italy,}\vskip 1mm
\centerline{\it and Universit\'a degli Studi di Milano,}\vskip 1mm
\centerline{\it Via Celoria, 16, 20133 Milano, Italy.}

\vskip 6mm
\centerline{\bf Abstract:} 
\vskip 7mm
\noindent 
In this paper I will discuss the possibility of a reionization of the 
Universe due to the photons emitted by evaporating primordial 
black holes (PBHs); this process should happen during the last 
stages of the PBHs life, when the particle emission is very intense.
I will study the time evolution of the ionization degree $x$, of 
the plasma temperature $T_{e}$ and of the photon number density $n_{\gamma}$
characterizing the Universe after the recombination epoch:
a system of coupled differential equations for these variables 
is solved in an analytical way, by assuming, as a photon source,
PBHs having an initial mass $M\sim 10^{14}$ g.
I will also take into account the PBH emission of 
quarks and gluons jets, which should be expected from 
PBHs having a temperature $T > \Lambda_{QCD}$.
The results I obtain prove that such a kind of reionization
is possible, being able to increase the ionization degree of the 
Universe from a value $x=0.002$ (just after the recombination) to 
values near 1 (when the black holes evaporation ends); in the same time,
the rise of the plasma temperature $T_{e}$ is limited by powerful 
cooling effects; therefore, such a reionization model does not predict
an excessive distortion of the Cosmic Microwave Background (CMB) spectrum,
in agreement with the experimental FIRAS
data on the comptonization parameter $y_{c}$ ($y_{c}<2.5 \times 10^{-5}$).
\vskip 7mm
\centerline{{\bf 1. PBH QUANTUM EVAPORATION AND PHOTON}} 
\centerline{{\bf EMISSION SPECTRUM.}}
\vskip 7mm
The possibility that the Universe was reionized after the recombination
is strongly suggested by many experimental evidences as, for instance,
the Gunn Peterson test $^{1}$.
The causes of such a reionization for the Universe are unclear
and, in general, a complete theory describing this 
phenomenon is not yet available.

In Refs. $^{2,}~^{3}$ I discussed a possible reionization mechanism 
based on the quantum evaporation of primordial black holes; 
why primordial? Due to the inverse proportionality 
existing between the mass and the temperature of an
evaporating black hole, we should expect a significant radiation 
emission only for small mass ($M << M_{\odot}$) BHs, therefore
formed immediately after the Big Bang.
The nature of the emitted particles obviously depends on the blackbody 
temperature of the PBHs: 
typically, BHs having a mass larger than $10^{17}~g$ emit massless particles 
only, like photons, neutrinos and, may be, gravitons.

In Refs. $^{4,}$ $^{5}$ Carr, Mac Gibbon and Webber showed that the quarks and 
gluons jets production should not be neglected when the BH mass falls 
below $10^{14}~g$ and the temperature $T$ becomes larger than the confinement 
scale $\Lambda_{QCD}$: in this case, the 
Hawking emission rate for particle production becomes $^{6}$:
$$
{dN_{x}\over dtdE}~=~\sum_{j}~\int^{+\infty}_{0}~{
\Gamma_{j}(Q,T) \over 2\pi\hbar }~\Big(~exp{Q\over T}\pm1~\Big)^{-1}~
{dg_{jx}(Q,E)\over dE}~dQ;\eqno(1.1)
$$
here 
the signs $\pm$ respectively refer to fermions and bosons;
$\Gamma$ is the absorption probability of the emitted species $^{7}$,
$x$ and $j$ respectively refer to the final and to the directly emitted 
particles and the last factor expresses
the number of particles with energy in the range $(E, E+dE)$ coming
from a jet having an energy equal to $Q$;
the fragmentation function $g_{jx}$ reads as follows $^{6}$:
$$
{dg_{jx}(Q,E)\over dE}~=~{1\over E}~\Bigg(~1-{E\over Q}~\Bigg)^{2m-1}~
\theta(E-km_{h}c^{2}),\eqno(1.2)
$$
where $m_{h}$ is the hadron mass, $k$ is a constant $O(1)$ and $m$ is an 
index equal to 1 for mesons and 2 for baryons. 
\vskip 7mm
\centerline{{\bf 2. THE TIME EVOLUTION OF THE IONIZATION DEGREE}}
\centerline{{\bf AND OF THE PLASMA TEMPERATURE FOR A REIONIZED UNIVERSE.}}
\vskip 7mm
As I discussed in Refs. $^{2}$ and $^{3}$, the basic equations that control the 
time evolution of the
ionization degree $x$ and of the plasma temperature $T_{e}$ are $^{8}$:
$$
{dx\over dt}~=~\ppi~+~\tco~-~\tri,\eqno(2.1)
$$
and 
$$
{d\tee\over dt}~=~-2~\scal~\tee~-~{\tee\over (1+x)}~{dx\over dt}
~+~{2\over 3(1+x)}~(\Gamma -\Lambda),\eqno(2.2)
$$
where $\ppi ,~\tco ,~\tri$ are, respectively, the photoionization, 
the collisional and the recombination rates. 
In eq. (2.2), the heating $\Gamma$ mainly comes from the 
photoionization process while the cooling $\Lambda$ takes into account 
the contributions due to the recombination, the collisional and excitation 
processes, the Compton scattering and to the 
expansion of the Universe: a detailed discussion of these effects can be found
in Ref. $^{3}$.

Both the equations (2.1) and (2.2) explicitly contain the photon 
number density $n_{\gamma}$, whose time evolution can be written as 
follows $^{2,}$ $^{3}$:
$$
{\parn\over\part}~+~\scal~{\parn\over\paro}~\omnor~-~2~\scal~
{\num (\omega, t)\over\omega}~\omnor~=
$$
$$
=~\Big(~{d\nure\over dt}
-{d\nupi\over dt}~\Big)~+~
{2\over \omega}~{d\omega\over dt}~[{\nupi - \nure}]~+~{\park\over
{\paro\part}}~\omnor;\eqno(2.3)
$$
here $\nupi,~\nure$ are the photon number densities respectively involved
in the processes of photoionization and recombination and $\omega_{NOR}$
is a normalization factor, equal to $10^{-6}$.
The last term in eq. (2.3) is the contribution of the photon source, 
in our case a number $n_{PBH}$ of evaporating primordial black holes.
A rough estimate of this parameter 
can be obtained by considering the following 
relations:
$$
\rho_{i}~\sim~{<M>~\nib\over R^{3}(t_{in})},\eqno(2.4)
$$
where $<M>\sim 10^{14}~g$ and
$$
R(t_{in})~\sim~R_{0}~(t_{0}/t_{in})^{\alpha};\eqno(2.5)
$$
here $R_{0}=1.25\times 10^{28}~cm~=1.4\times 10^{10}~lyr$ in a typical
cosmological model $^{9}$ and $\alpha =-0.5$.
Then, if one assumes for the PBHs density a behaviour that approximately 
scales as a power with exponent $2/3$ of the time, 
the present density parameter $\Omega_{PBH}$ varies in a range:
$$
\Omega_{PBH}~=~ {\rho_{0PBH} \over \rho_{cr}}~=~
1.12\times 10^{-12}~\div 1.65\times 10^{-8};
\eqno(2.6)
$$
this rough estimate reproduces quite well the present experimental upper 
limit for $\Omega_{PBH}$, coming from the CMB constraints $^{4}$:
$
\Omega_{PBH}\leq
(7.6\pm 2.6)\times 10^{-9}~h^{(-1.95\pm 0.15)}.
$
\vskip 7mm
\centerline{{\bf 3. THE NUMERICAL SOLUTION OF THE DIFFERENTIAL}}
\centerline{{\bf EQUATION SYSTEM: RESULTS AND CONCLUSIONS}}
\vskip 7mm
A detailed discussion of the analytical solution of eqs. (2.1), (2.2) and (2.3)
can be found in Refs. $^{2,}$ $^{3}$; here I want only to recall the main 
results coming from this analysis.

In figs. 1 and 2 I plotted the behaviour of the ionization degree $x$ 
and of the plasma temperature $T_{e}$ for a reionization redshift $\zre=30$:
as one can see, the BH-induced 
reionization of the Universe is indeed possible but it is partial only; 
a quite relevant effect is obtained for an evaporation redshift corresponding 
to $\zre \leq 30$, while for higher values of $\zre$ the process
of PBHs quantum evaporation cannot produce an appreciable 
phenomenon of reionization.

The plasma heating level is 
a fundamental test for all the reionization models: in fact, an excesssive
heating would contradict the FIRAS upper limit 
on the comptonization parameter $y_{c}$ ($y_{c}<2.5 \times 10^{-5}$).
In our case, the plasma heating is limited by a powerful
cooling: in eq. (2.1) the $\ppi$ term is suppressed and the $\tri$ term 
is enhanced $^{3}$.

Finally, in this model, we have none significant suppression of
the CMB temperature fluctuations on small angular scales:
Bond and Efstathiou $^{10}$ and Vittorio and Silk $^{11}$ suggest
that a reionization at $z_{R}\sim 50$ could indeed suppress 
these fluctuations that, following the predictions 
of the CDM models and texture scenarios $^{12}$, would result too large; but,
as I told, in my model the reionization for $z_{R} >30$ is uneffective.

The PBHs formation time should be put very far in the past:
considering the jet emission contribution in the photon spectrum,
I can obtain a well
balanced reionization process (i.e. a high ionization degree without 
an excessive plasma heating) only for a formation time
contemporary to the Big Bang and for an initial density 
$\rho_{i}=10^{17}~\div~10^{24}g/cm^{3}$, corresponding to a 
present density parameter $\Omega_{PBH}\sim 1.12\times 10^{-12}$ 
$\div~1.65\times 10^{-8}$.

The behaviour of $x$ can be approximated by an exponential
function: this result justifies my analysis of Ref. $^{13}$,
where I studied the consequences of an exponential reionization 
of the Universe on the polarization of the Cosmic Microwave Background. 

\vskip 1cm
\centerline{{\bf ACKNOWLEDGEMENTS}}
\vskip 7mm
I would like to address a sincere 
thank to the Universit\'a degli Studi di Pavia for its financial
support and to the people working there, for many interesting discussions
on these arguments. I am also grateful to all the people of the 
Astrophysics Section of the University of Milano; in particular my affectionate
thank goes to my friends Emma and Francesco.
\vskip 1cm
\centerline {\bf{REFERENCES}}
\vskip 7mm

\item{[1]} Gunn, J.E., Peterson, B.A., Astroph. J.,{\bf 142}, (1965), 1633.

\item{[2]} Gibilisco, M: ''Reionization of the Universe induced by 
Primordial Black Holes'', accepted for pub. in 
Int. Journ. of Mod. Phys A, In press.

\item{[3]} Gibilisco, M: ``The influence of quarks and gluons jets 
coming from PBH on the reionization of the Universe'',
preprint babbage astro-ph/9604116,
accepted for pub. in Int. Journ. of Mod. Phys A, Sept. 1996.

\item{[4]} Mac Gibbon, J.H., Carr, B.J., Astroph. J., {\bf 371}, (1991), 447;

\item {} Mac Gibbon, J.H., Webber, B.R., Phys. Rev., {\bf D41}, (1990), 3052.
                                         
\item{[5]} Mac Gibbon, J. H., Phys. Rev., {\bf D44}, (1991), 376.

\item{[6]} Carr, B. J., Astronomical and Astroph. Transactions, Vol. {\bf 5},
(1994), 43.

\item{[7]} Page, D. N., Phys. Rev., {\bf D13}, (1976), 198.

\item{[8]} Durrer, R., Infrared Phys. Technol., {\bf 35}, (1994), 83.

\item{[9]} Misner, C.W., Thorne, K.S., Wheeler, J.A.: ``Gravitation'', (1973),
p. 738, W. H. Freeman and Co. Eds., San Francisco.

\item{[10]} Bond, J.R., Efstathiou, G., Astroph. J.,{\bf 285}, (1984), L45.

\item{[11]} Vittorio, N., Silk, J., Astroph. J.,{\bf 285}, (1984), L39.

\item {[12]} Turok, N., Phys. Rev. Lett., {\bf 63}, (1989), 2625;
                        
\item{} Turok, N., Spergel, D.N., Phys. Rev. Lett., {\bf 64}, (1990), 2736;

\item{} Durrer, R., Phys. Rev., {\bf D42}, (1990), 2533.

\item {} Tegmark, M., Silk, J., ''On the inevitability of Reionization:
Implications for Cosmic Microwave Background Fluctuations'', Preprint
babbage astro-ph/9307017, June 1993.

\item{[13]} Gibilisco, M., Intern. Journal of Modern Phys, {\bf 10A}, (1995)
3605.

\bye